\documentclass[10pt]{article}

\usepackage{bbold}
\usepackage{amssymb}
\usepackage{graphics}

\def\d{\partial}
\def\m{\mu}
\def\n{\nu}

\def\be{\begin{equation}}
\def\ee{\end{equation}}
\def\beq{\begin{equation}}
\def\eeq{\end{equation}}
\def\bea{\begin{eqnarray}}
\def\eea{\end{eqnarray}} 
\def\beqa{\begin{equation}\begin{array}{l}}
\def\eeqa{\end{array}\end{equation}}

\def\eqref#1{eq.~(\ref{eq:#1})}

\begin{document}

\thispagestyle{empty}

\vspace{.8cm}
\setcounter{footnote}{0}
\begin{center}
{\large{\bf 
Acausality of Massive Charged Spin~2 Fields
    }}\\[3mm]

S, Deser$^\clubsuit$
and A. Waldron$^{\spadesuit}$
\\[3mm]

{\em\small  
$^\clubsuit\;$Physics Department, Brandeis University, Waltham,
MA 02454, USA\\ 
{\tt deser@brandeis.edu}}\\[3mm]
{\em\small
$^\spadesuit\;$Mathematics Department, University of California, Davis,
CA 95616, USA\\ 
{\tt wally@math.ucdavis.edu}}\\[7mm]

{\sc Abstract}\\
\end{center}

{\small
\begin{quote}

The recent claim in {\tt hep-th/0302225} that, contrary to all previous
work, massive charged $s=2$ fields propagate causally is false.

\bigskip

\end{quote}
}






\noindent
Charged massive 
spin~2 fields
are known to propagate 
acausally~\cite{Kobayashi,Capri,Deser}
because the constraints 
fail to prevent superluminal excitations for all values of
the background E/M field. Alternatively, this phenomenon may be viewed
as a breakdown of the constraints at these points in field
space~\cite{Capri,Deser}.
A recent paper~\cite{Novello} claims that there is no acausality, based on 
computations involving ``Fierz tensors'' $F_{\m\n\rho}=-F_{\n\m\rho}$
built from a certain combination of first derivatives of the usual spin~2
field $\varphi_{\m\n}$. We shall point out an error in these computations.
The usual method of characteristics is followed in which the spin~2
field has a leading discontinuity at a shock 
$[\d_\mu \d_\nu\varphi_{\rho\sigma}]=k_\mu k_\nu \epsilon_{\rho\sigma}$.
The authors introduce the quantity $F_\mu\equiv D_\mu \varphi - D.\varphi_\mu$
whose derivative has discontinuity
$[\d_\mu F_\nu]=k_\mu k_\nu \epsilon_\rho^\rho 
- k_\mu k^\rho\epsilon_{\nu\rho}\equiv k_\mu f_\nu$.
They give a correct derivation, based on discontinuities of the constraint
equations, of the relations $f_\mu f^\mu=k_\mu f^\mu=0$. However, they then try 
to conclude that the normal vector satisfies the lightlike condition $k_\mu k^\mu=0$ 
which would imply causal propagation. Unfortunately, this neglects the fact that 
$\varphi_{\mu\nu}$ is a charged and therefore complex field. Since $f_\mu$ is a complex vector, 
the conclusion that $k_\mu$ is lightlike is clearly erroneous. Indeed, as demonstrated 
in~\cite{Kobayashi,Capri,Deser}, there are acausal solutions to this system that
satisfy $f_\mu f^\mu=k_\mu f^\mu=0$. [The authors additionally claim that the
massless gauge invariance that appears in the eikonal limit can be used to gauge away acausalities.
This is patently false since the underlying massive system is not gauge invariant.]
We (re)conclude that charged massive $s=2$ shares its $s=3/2$
counterpart's classic propagation problems~\cite{Johnson,Velo} in 
external fields,
irrespective of the formalism employed. This work was supported by the 
National Science Foundation under grants PHY99-73935 and PHY01-40365.

\end{document}